\documentclass[aps,pre,onecolumn,showpacs,amsmath,amssymb,longbibliography,notitlepage,11pt,tightenlines]{revtex4-1} 
\usepackage{graphicx} 
\usepackage{dcolumn} 
\usepackage{bm} 

\RequirePackage{lettrine} 
\newcommand{\dropcapfont}{\fontfamily{lmss}\bfseries\fontsize{26pt}{28pt}\selectfont}
\newcommand{\dropcap}[1]{\lettrine[lines=2,lraise=0.05,findent=0.1em, nindent=0em]{{\dropcapfont{#1}}}{}}

\begin{document}

\title{Hyperuniformity with no fine tuning in sheared sedimenting suspensions}

\author{Jikai Wang}
\author{J. M. Schwarz}
\author{Joseph D. Paulsen}
\email{jdpaulse@syr.edu}
\affiliation{Department of Physics and Soft and Living Matter Program, Syracuse University, Syracuse, NY 13244, USA}

\date{\today}

\begin{abstract}
Particle suspensions, present in many natural and industrial settings, typically contain aggregates or other microstructures that can complicate macroscopic flow behaviors and damage processing equipment. 
Recent work found that applying uniform periodic shear near a critical transition can reduce fluctuations in the particle concentration across all length scales, leading to a hyperuniform state. 
However, this strategy for homogenization requires fine tuning of the strain amplitude. 
Here we show that in a model of sedimenting particles under periodic shear, there is a well-defined regime at low sedimentation speed where hyperuniform scaling automatically occurs. 
Our simulations and theoretical arguments show that the homogenization extends up to a finite lengthscale that diverges as the sedimentation speed approaches zero. 
\end{abstract}

\maketitle

\bigskip
\dropcap{P}article suspensions can respond to flow in dramatic ways. 
Steady shear can cause their viscosity to jump by orders of magnitude in some situations, or to plummet in others~\cite{Wagner09,Brown10,Cheng11}. 
Interparticle or external forces such as gravity can alter suspension properties over time~\cite{Fielding00}. 
These effects put large demands on handling and processing. 
Thus, methods are desired for obtaining homogeneous particle distributions with predictable mechanical properties, as a platform for further handling. 
On small lengthscales, one wants to break up aggregates or pockets of high concentration, since particles moving in close proximity cause significant dissipation. 
On large lengthscales, one wants different parts of the sample to have similar particle concentrations so that the rheological response is stable and reliable. 

\begin{centering}
\begin{figure}[tb]
  \includegraphics[width=8.3cm]{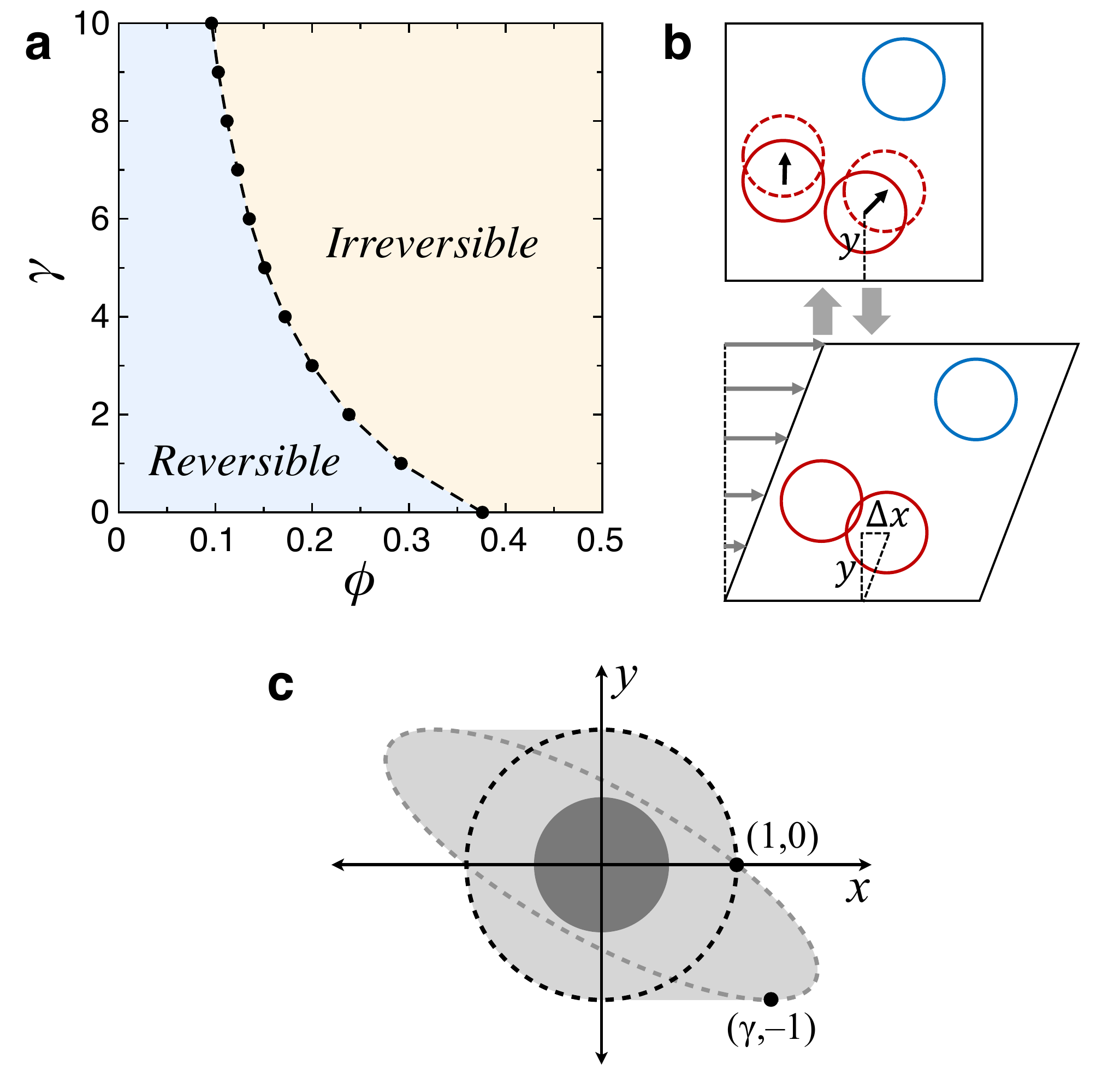}
  \caption{
  \textbf{Sheared non-Brownian suspension model after Ref.~\citealp{Corte08}.} 
  \textbf{(a)} Phase diagram showing reversible steady states at low concentration, $\phi$ and strain amplitude, $\gamma$. 
  Outside this region, a finite fraction of particles collide during each cycle in the steady state. 
  Dashed line: Critical phase boundary. 
  Data show the largest $\gamma$ where we obtain a reversible state in simulations with $L=200d$. 
  \textbf{(b)} Simulation algorithm. In each shear cycle, particles are displaced a horizontal distance $ \Delta x = \gamma y$ and then returned to their initial positions. Particles that overlap (red) are given random kicks, to simulate local irreversibility due to collisions.
  \textbf{(c)} Interaction region around a particle. 
  A second particle with its center anywhere inside the dashed circle will overlap with the particle at the origin (shown as a dark circle); they would both receive a random kick. 
  Shearing the system expands the interaction region to the entire shaded area (shown for one value of $\gamma$), which contains the points that are covered by continuously shearing the  dashed circle up to strain amplitude $\gamma$ and back. 
  }
  \label{model}
\end{figure}
\end{centering}

Recent experiments have shown that non-Brownian suspensions can be driven to well-behaved states simply by applying cyclic, low-Reynolds number shear from the boundaries~\cite{Pine05,Corte08,Paulsen14,Pham15,Schrenk15}. 
For small strain amplitudes $\gamma$, the particles automatically self-organize into reversible steady states, whereas 
for amplitudes larger than a critical value, $\gamma_c$, the particles follow irreversible paths indefinitely. 
An underlying non-equilibrium phase transition has been rationalized by simulations with simple particle kinematics~\cite{Corte08} (see the phase diagram in Fig.~\ref{model}a), and the transition has been shown to directly affect the rheological response in experiments~\cite{Corte08,Franceschini11,Paulsen14}. 
Further simulations suggest that in such suspensions, the particles should exhibit extremely uniform spatial distributions when driven for many cycles at the critical strain amplitude, $\gamma_c$~\cite{Hexner15,Tjhung15}. 
These distributions are called `hyperuniform' and are characterized by density fluctuations that decay rapidly as one looks over larger and larger lengthscales~\cite{Torquato03,Jack15,Ma17}. 
Shearing at $\gamma_c$ is thus an attractive method for homogenizing a suspension. 
Yet, from a practical standpoint it is hindered by requiring precise tuning of the strain amplitude~\cite{Hexner15}. 

Here we present a robust method for obtaining a hyperuniform state in a viscous suspension. 
Based on recent work by Corte \textit{et al.},~\cite{Corte09}, we introduce a small density mismatch between the suspending fluid and the particles so that they sediment slowly under gravity. 
In this situation, cyclic shear was found to re-suspend the particles up to a height where they achieve the critical concentration, $\phi_c$~\cite{Corte09}. 
Our simulations and theoretical arguments show that there is a well-defined regime at low sedimentation speed where this combination of sedimentation and shear serves to homogenize the system. 
In this regime, density fluctuations are significantly suppressed up to a finite lengthscale. 
We show that this lengthscale is set by small vertical gradients in the particle concentration, and it can be made arbitrarily large simply by slowing the sedimentation rate. 
We thereby construct a phase diagram for this `self-organized hyperuniformity', which is in good agreement with our simulation results.

\bigskip
\noindent\textbf{\large Results}\\
\noindent\textbf{Simulations.}
We use a simulation model originally developed by Refs.~\cite{Corte08,Corte09}. 
This method captures a wide range of behaviors seen in experiments on sheared non-Brownian suspensions, including self-organization~\cite{Corte08} and novel memory effects~\cite{Keim11,Paulsen14}. 
We place $N$ particles of diameter $d$ in a square box of width $L$ with area fraction $\phi = N \pi (d/2)^2 / L^2$. 
The box is periodic on the left and right sides, and the top and bottom are hard walls. 
We use units where $d=1$ so that lengths are in particle diameters, and we measure time in units of cycles. 

Following Ref.~\cite{Corte09}, each cycle consists of particle sedimentation and shear. 
First, all particles sediment vertically a distance $v_\text{s}$. 
Shear is then applied in several steps as illustrated in Fig.~\ref{model}b. 
First, particles are displaced with an affine transformation, $(\Delta x, \Delta y) = (\gamma y,0)$, where $y$ is the distance from the particle center to the bottom wall. 
All particles are then returned to their original unsheared positions. 
Particles that overlap during this transformation are given a kick in a random direction with a magnitude chosen uniformly between $0$ and $\epsilon$, where $\epsilon=0.5$ except where otherwise stated. 
The effect of the shearing is to make particles collide that are within an interaction region, like the one sketched in Fig.~\ref{model}c. 
(In previous studies, varying the kick size or collision kinematics did not change the qualitative results~\cite{Keim13}.)

\begin{centering}
\begin{figure*}[t]
  \includegraphics[width=15.5cm]{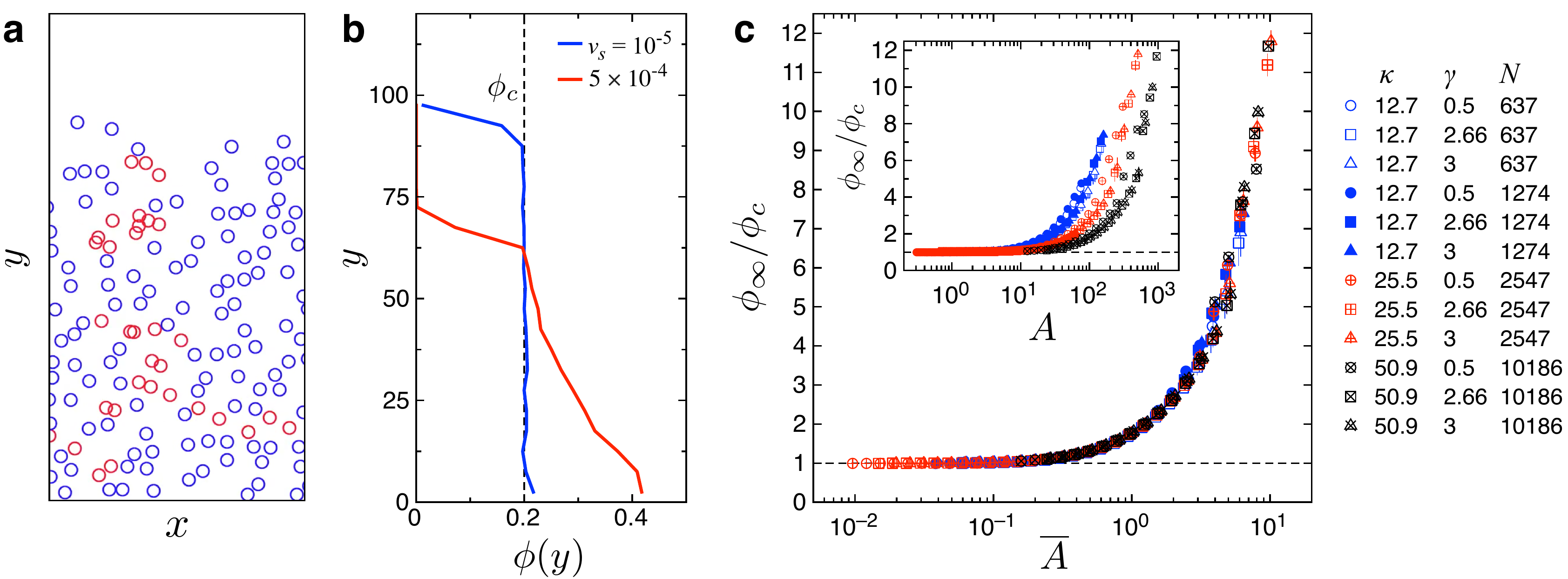}
  \caption{
  \textbf{Self-organized criticality at low sedimentation speed.} 
  \textbf{(a)} Snapshot of a system in the steady state. 
  Red particles are colliding in the current cycle. 
  \textbf{(b)} Particle concentration (plotted on the $x$ axis) versus vertical coordinate, $y$, at low and high sedimentation velocity $v_s$, with $N=2547$, $\gamma=3.0$, $\kappa=25.5$, and $L=100$. 
  Vertical dashed line shows $\phi_c(\gamma=3)=0.20$. 
  \textbf{(c)} Scaled steady-state concentration $\phi_\infty/\phi_c$, measured over a wide range of system parameters ($\kappa$, $\gamma$, and $N$ as shown in legend) and velocities ($10^{-5} < v_s < 10^{-2}$). 
  \textit{Inset:} Measurements versus the parameter $A=(\pi/\phi_{c})^{3/2} d^3 \kappa^2 v_s / 32D$, proposed by Ref.~\cite{Corte09}. 
  The data are not collapsed. 
  \textit{Main:} The data collapse when replotted versus $\overline{A}$ (Eq.~\ref{ourA}). 
  For $\overline{A}\ll1$ (low sedimentation speed), the steady-state concentration $\phi_\infty$ is equal to the critical value $\phi_c$. 
  Error bars show the size of fluctuations within the steady state. 
  }
  \label{Phi_A}
\end{figure*}
\end{centering}

\medskip
\noindent\textbf{Self-organized criticality.}
Cort\'e et al.~\cite{Corte09} recently showed that for sufficiently slow sedimentation a critical state is automatically reached, offering a rare example of self-organized criticality seen in both simulation and experiment~\cite{Bak87}. 
This behavior occurs when the steady-state concentration of the particles is equal to the critical concentration, $\phi_c$, and it can be anticipated from simple arguments. 
At any $v_s$, the particles settle to a steady-state height where sedimentation and diffusion balance as pictured in Fig.~\ref{Phi_A}a (see also Supplementary Videos 1 and 2), in a process called viscous resuspension~\cite{Acrivos93}. 
For slower sedimentation, this balance leads to a higher suspension height, as shown by the vertical concentration profiles in Fig.~\ref{Phi_A}b. 
Hence, at lower $v_s$, the average concentration throughout the suspension in the steady state, $\phi_\infty$, is also lower (where the subscript indicates this is the steady-state value). 
Crucially, because the diffusion process is driven by collisions, the particles stop spreading apart when they are just far enough away to stop colliding, so the concentration cannot decrease below $\phi_c$. 
Thus, $\phi_\infty \rightarrow \phi_c$ as $v_s \rightarrow 0$.

Quantitatively, the critical concentration is achieved when a suitably-chosen sedimentation timescale, $\tau_s$, is much larger than a diffusion timescale, $\tau_D$. 
Cort\'e et al.~\cite{Corte09} proposed that $\tau_s$ is set by the time to sediment a mean particle spacing (a distance $\sqrt{\pi d^2 / 4\phi}$), and $\tau_D$ is the characteristic time for a particle to diffuse over the total height of the suspension (i.e., $\tau_D = h^2/4D$ where $D$ is the coefficient of diffusion for a non-sedimenting system at $\phi=2\phi_{c}$, and $h = \pi d^2 \kappa / 4 \phi$ is the suspension height with $\kappa = N/L$ being the linear density of particles along the $x$ axis). 
In a critical state where $\phi=\phi_c$, the ratio of these timescales is: $A=\tau_D/\tau_s=(\pi/\phi_{c})^{3/2} d^3 \kappa^2 v_s / 32D$. 

The inset to Fig.~\ref{Phi_A}c shows our measurements of $\phi_\infty/\phi_c$, where we vary velocity $v_s$, linear density $\kappa$, strain amplitude $\gamma$, and system size $N$ over a broad range. 
The data indeed approach $1$ for small $A$, but they are clearly not collapsed. 
(Reference~\cite{Corte09} set their expression for $A$ to be $8$ times this value; this merely shifts all the data along the $x$ axis by a fixed amount.) 

We propose that the timescales for diffusion and sedimentation should instead be considered over the same lengthscale. 
Taking $\tau_D$ and $\tau_s$ as the timescales for particle transport over the critical height of the bed of particles, $h_c$, we obtain:
\begin{equation}
  \overline{A}=\frac{\tau_D}{\tau_s}=\frac{\pi}{16}\frac{d^{2}\kappa v_s}{\phi_c D},  
  \label{ourA}
\end{equation}

\noindent which serves as a non-dimensional sedimentation speed. 
This expression produces an excellent collapse of the data, as shown in Fig.~\ref{Phi_A}c. 
(The same expression also collapses the data in a version of the algorithm where a separate kick is given for each particle encountered in a cycle; see Supplementary Note 1 and Supplementary Fig.~1.)

\begin{figure*}[t]
  \includegraphics[width=14.5cm]{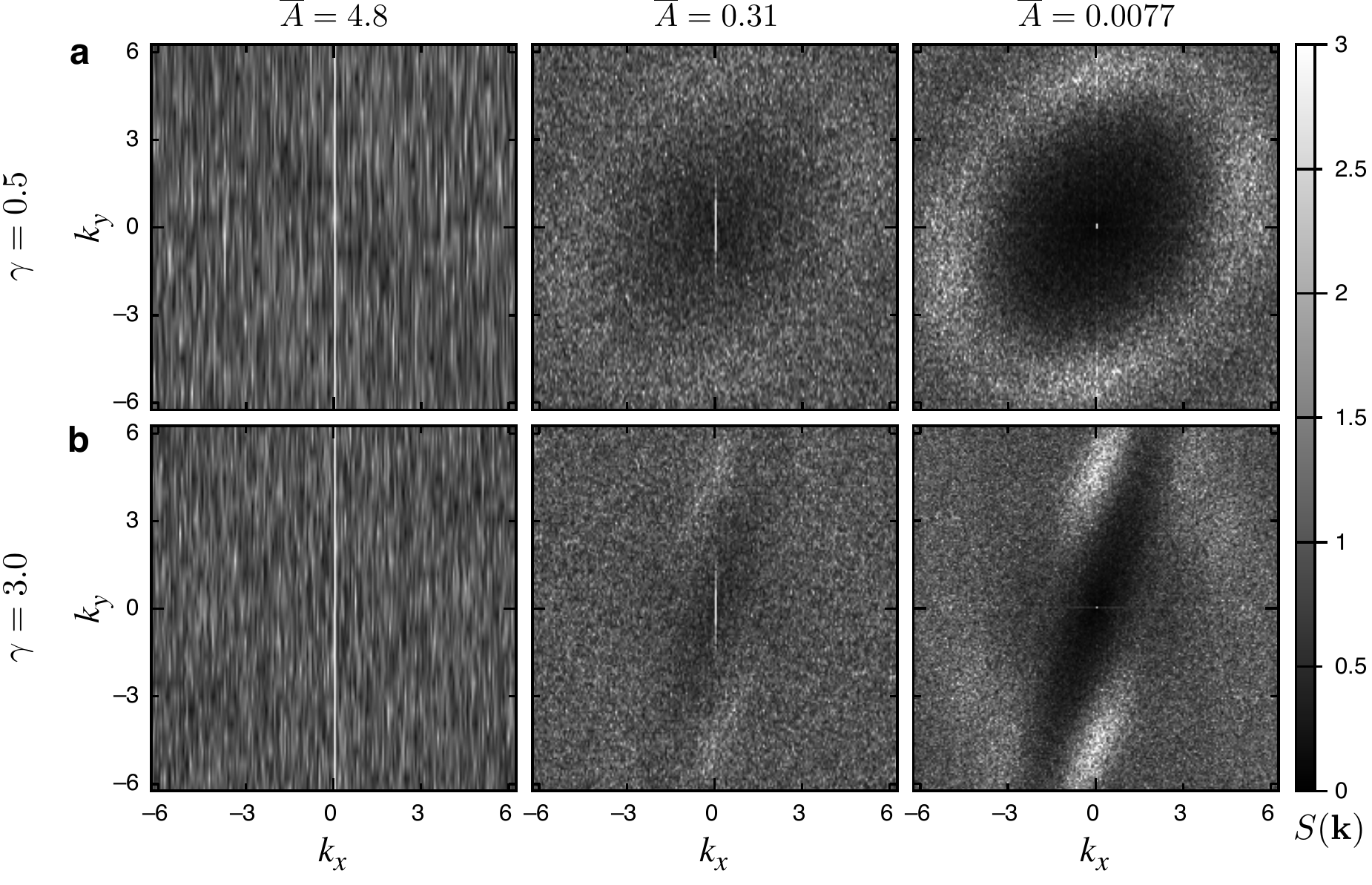}
  \caption{
  \textbf{Static structure factor.} 
  $S(\mathbf{k})$ measured in 2D simulations with $N=4827$ and $\kappa=34.1$ at three values of $\overline{A}$. 
  Each panel is an average of $5$ systems analyzed at a single snapshot in time in the steady state. 
  Values are calculated using Eq.~\ref{sofk} at wavevectors $k_x$ and $k_y$ where an integer number of wavelengths fill the region populated by the particles. 
  We avoid the diffuse boundary layer at the top of the sample by staying in the bottom 99\% of the particles. 
  \textbf{(a)} Results for $\gamma=0.5$. 
  At low $\overline{A}$ the structure factor is vanishing near the origin, signaling hyperuniformity. 
  (The pattern is not radially symmetric because of the anisotropic driving.) 
  \textbf{(b)} Results for $\gamma=3$. 
  Despite larger anisotropy in the structure, hyperuniformity occurs at low $\overline{A}$. 
  }
  \label{structure_factor}
\end{figure*}

\medskip
\noindent\textbf{Structure factor.} 
At low sedimentation speed, the suspension is in a critical state characterized by a power-law distribution of avalanches that are set off by individual collisions~\cite{Corte09}. 
Although a suggestive connection has been identified between criticality and hyperuniformity~\cite{Hexner15}, there is presently no deductive link. 
To see whether hyperuniformity can survive the dynamics of continual resuspension, we now look for it in our simulations. 
Following previous studies~\cite{Dreyfus15, Hexner15, Tjhung15, Weijs15, Schrenk15}, we consider the structure factor defined by: 
\begin{equation}
S(\mathbf{k}) = \frac{1}{N} \Big| \sum_j e^{i \mathbf{k} \cdot \mathbf{r}_j } \Big|^2 , 
\label{sofk}
\end{equation}

\noindent where $\mathbf{k}$ is a two-dimensional (2D) wavevector and $\mathbf{r}_j$ is the location of the $j^\text{th}$ particle center. 
Density fluctuations over long distances in real space affect $S(\mathbf{k})$ near the origin in reciprocal space; the hallmark of hyperuniformity is that $S(\mathbf{k}) \rightarrow 0$ as $\mathbf{k} \rightarrow 0$. 

Figure~\ref{structure_factor} shows our measurements of the structure factor for two values of the strain amplitude, $\gamma$, and three values of the non-dimensional sedimentation speed, $\overline{A}$. 
For large sedimentation speeds, $\overline{A} > 1$, the data are featureless and show that density fluctuations exist on all lengthscales. 
(The thin white band is due to vertical concentration gradients.) 
At smaller $\overline{A}$, the values decrease near the origin. 
Hyperuniformity is clearly present at $\overline{A}=0.0077$.  
Crucially, each strain amplitude produces a hyperuniform state with no fine tuning of the driving.

\begin{centering}
\begin{figure*}[t]
  \includegraphics[width=16.4cm]{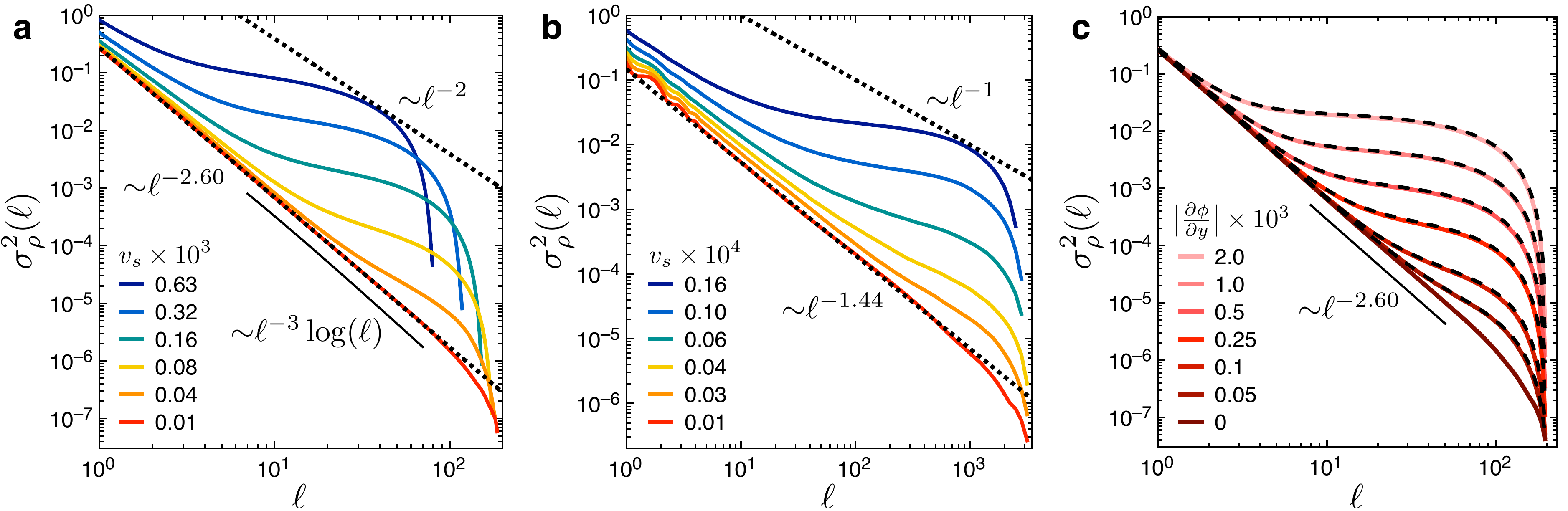}
  \caption{
  \textbf{Number density fluctuations.}
  \textbf{(a)} 
  Steady-state variance of the number density, $\sigma_\rho^2(\ell)$, versus window size, $\ell$, in 2D simulations with sedimentation and shear. 
  Each point is an average of 50 measurements with $N=9677$, $\gamma=3.0$, and $\kappa=48.4$. 
  At low sedimentation speed, hyperuniform density fluctuations are observed, as shown by the dotted line following $0.27 \ell^{-2.60}$. 
  An uncorrelated system would scale as $\ell^{-2}$. 
  \textbf{(b)} 
  Analogous results in 1D simulations, averaged over 20 systems with $N = 3000$ and $\gamma = 1$. 
  At low sedimentation speed, hyperuniform density fluctuations are observed over 3 decades in $\ell$, following $0.15 \ell^{-1.44}$. 
  An uncorrelated system would scale as $\ell^{-1}$. 
  \textbf{(c)} Variance of the number density for initially hyperuniform 2D systems that were scaled along the $y$ axis to impose a constant concentration gradient, $|\partial \phi/\partial y|$. 
  Each curve is an average over 31 systems with $N=10186$, $\gamma=3.0$, $L=200$, and $\phi=0.2$. 
  The variance increases with the size of the gradient. 
  The data are captured by summing the measurement at zero gradient with a term due to a uniform gradient that we calculate in the continuum limit (dashed lines: Eq.~\ref{decomposition}, using Eq.~\ref{var_grad} with the value of $|\partial \phi/\partial y|$ in the legend). 
  }
  \label{variance-fig}
\end{figure*}
\end{centering}

\medskip
\noindent\textbf{Density fluctuations in real space.}  
To probe the system further, we investigate density fluctuations in real space. 
We consider the number density of particles, $\rho$, within circular regions of diameter $\ell$, centered at random locations within the bed of particles, where we avoid the diffuse boundary layer at the top of the sample by staying in the bottom 99\% of the particles. 
Denoting the variance of the number density over these samples by $\sigma_{\rho}^2(\ell) \equiv \langle \rho^2(\ell) \rangle - \langle \rho(\ell) \rangle^2$, hyperuniformity is characterized by the rate of decay of $\sigma_{\rho}^2(\ell)$ with respect to window size:
\begin{equation}
  \sigma_{\rho}^2(\ell) \propto \ell^{-\lambda}, 
  \label{def}
\end{equation}

\noindent with $\lambda$ exceeding the spatial dimension of the system. 

Figure~\ref{variance-fig}a shows $\sigma_{\rho}^2(\ell)$ for several sedimentation speeds, $v_s$. 
At the lowest $v_s$ we observe hyperuniform behavior with a scaling exponent of $\lambda=2.60\pm0.04$. 
(The data are also consistent with $\sigma_{\rho}^2(\ell) \sim \ell^{-3}\log(\ell)$, a scaling that occurs in jammed packings~\cite{Donev05,Zachary11} and in an isotropic version of the sheared-suspension model in the presence of noise~\cite{Hexner17a}.) 
Notably, our measurements at low velocity show the same scaling as our simulations of non-sedimenting particles at $\phi=\phi_c$ (bottom data in Fig.~\ref{variance-fig}c with $\lambda=2.60$). 
At larger velocities the variance $\sigma_{\rho}^2(\ell)$ gradually increases, which is noticeable first at long lengthscales and then at smaller and smaller $\ell$. 

The tails at large $\ell$ are due to the finite size of the system.  
In particular, when the window size $\ell$ is comparable to the height of the suspension bed, the sampling windows are forced to overlap. 
This reduces the variation in number density as the measurements are not independent. 
This effect was identified in Ref.~\cite{Dreyfus15}; we investigate it further in Supplementary Note 2 and Supplementary Fig.~2. 
The steady-state height of the bed is also shorter at larger velocities, which limits the data to a smaller range of $\ell$. 

To demonstrate hyperuniform scaling of the variance up to even larger lengthscales, we model an analogous system in one dimension (1D). 
The system is oriented vertically and sedimentation is applied as in the 2D case (see Supplementary Video 3). 
Following previous work on such 1D models~\cite{Corte08,Paulsen17}, any particle that is within an interaction distance $\gamma=1$ of another receives a random kick with a magnitude between $0$ and $\epsilon$, displacing it either up or down with equal probability. 
Figure~\ref{variance-fig}b shows $\sigma_{\rho}^2(\ell)$ measured in the steady state for different sedimentation speeds. 
At low $v_s$, hyperuniform scaling occurs over three decades in length with $\lambda = 1.44 \pm 0.02$.

\medskip
\noindent\textbf{Loss of hyperuniform scaling by concentration gradients.} 
In both 2D and 1D, the loss of hyperuniform scaling at high sedimentation speeds is no surprise---these systems show large vertical concentration gradients at high $v_s$, as shown already in Fig.~\ref{Phi_A}b and seen in earlier work on this model~\cite{Corte09}. 
Nevertheless, one wants to know how small the velocity must be to prepare a system in a hyperuniform state. 
In the remainder of this article, we build up a general quantitative framework that answers this question. 
Our approach is to split the total variance of the number density into two additive terms: one from the statistics of the particles in a critical state, $\sigma_\rho^2(\ell)_c$, and the other capturing the effect of a global concentration gradient. 
That is, 
\begin{equation}
  \sigma_\rho^2(\ell)_\text{total} = \sigma_\rho^2(\ell)_c + \sigma_\rho^{2}(\ell)_\text{grad}. 
  \label{decomposition}
\end{equation}
Our main task is to establish a quantitative description of $\sigma_\rho^{2}(\ell)_\text{grad}$. 
\noindent As a result of this analysis, we establish a finite lengthscale $\ell_\text{H}$ beyond which hyperuniform scaling is lost. 

To begin, we study the effect of system-spanning concentration gradients on number density fluctuations in a well-controlled setting. 
First, we generate hyperuniform distributions of particles by shearing a non-sedimenting system at $\gamma \approx \gamma_c$ until it reaches a reversible steady state. 
We then adjust the $y$ positions of these particles to create a uniform vertical concentration gradient. 
(The $y$-coordinate map is uniquely determined by requiring that the particle concentration maps as $\phi_0 \rightarrow \phi(y) = \phi_0 + |\partial \phi/\partial y|(h - 2y) / 2$ for a continuum system with initially uniform concentration $\phi_0$; see Supplementary Note 3 and Supplementary Fig.~3 for details.) 
Figure~\ref{variance-fig}c shows the variance $\sigma_{\rho}^2(\ell)$ of these distorted systems for different values of the gradient, $|\partial \phi/\partial y|$. 
As in the full sedimentation simulations, the variance is noticeably affected at large $\ell$ for small perturbations and then at shorter and shorter lengthscales as the perturbation size increases. 

We can understand these variance curves from simple arguments. 
We calculate the variance of the concentration for a continuous field $\phi(y)$ with a uniform vertical gradient (i.e., $\partial \phi/\partial y = \text{const.}$) in a rectangular domain of height $h$ and width $L$. 
We consider the case with periodic boundary conditions on the left and right sides to match our simulations. 
The concentration at position $(x,y)$ is given by: $\phi(x,y) = \phi_0 + \frac{1}{2}(\phi_b-\phi_t)(1 - 2y/h)$, where $\phi_b$ and $\phi_t$ are the concentrations at the bottom and top of the domain, and $\phi_0 = (\phi_b+\phi_t)/2$ is the mean concentration. 
The variance of the concentration is given by: $\sigma_\phi^{2}(\ell)=\int[\phi(x,y)-\phi_0]^{2}f(x,y) dx dy$, where $f(x,y)=1/[L(h-\ell)]$, reflecting the fact that the sampling window cannot cross the top or bottom of the domain (i.e., $\ell/2 < y < h-\ell/2$). 
Computing this variance and converting from concentration to number density, we find:
\begin{equation}
  \sigma_\rho^{2}(\ell)_\text{grad} = \frac{4}{3 \pi^2} \left( \frac{\partial \phi}{\partial y} \right)^2 (h-\ell)^2, 
  \label{var_grad}
\end{equation}

\noindent where $\partial \phi / \partial y = (\phi_t-\phi_b)/h$. 
The total variance in the discrete particle system is obtained by adding this result to the variance of the corresponding system with no concentration gradient (i.e., $\partial \phi / \partial y= 0$), as anticipated by Eq.~\ref{decomposition}. 
Figure~\ref{variance-fig}c shows that this prediction is in excellent agreement with the data across all lengthscales and over a large range of gradients. 

The similarity between Fig.~\ref{variance-fig}a,b for sedimentation simulations and Fig.~\ref{variance-fig}c for the effect of a simple linear distortion is striking. 
This result suggests that the density fluctuations in this system can be largely accounted for by understanding these gradients. 
We now move to quantify the strength of the vertical concentration gradients that arise in the model.

\begin{centering}
\begin{figure}[t]
  \includegraphics[width=7.5cm]{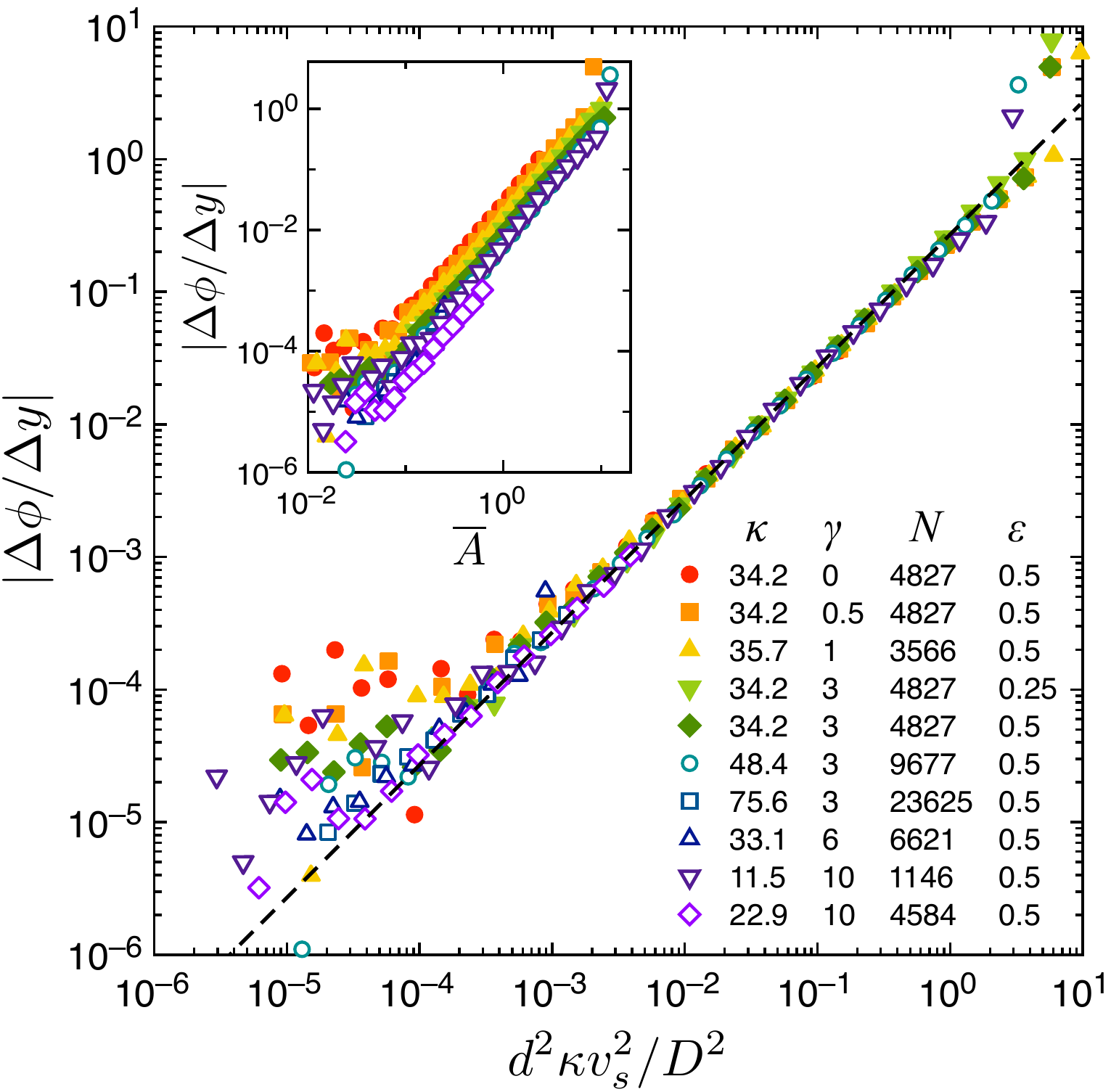}
  \caption{
  \textbf{Steady-state vertical concentration gradient.} 
  \textit{Inset:} $|\Delta \phi/\Delta y|$ versus $\overline{A}$, obtained by averaging over 5 systems. 
  \textit{Main:} The data are collapsed when plotted versus $d^{2}\kappa v_{s}^{2}/D^{2}$. 
  Dashed line: Scaling argument with fitted numerical prefactor, Eq.~\ref{x_coordinate_slope}. 
  }
  \label{slope-yprofile}
\end{figure}
\end{centering}

\medskip
\noindent\textbf{Magnitude of vertical concentration gradient.} 
We measure the mean steady-state concentration gradient, $|\Delta\phi/\Delta y|$, by fitting a straight line to the concentration profile, where we fit only the middle $60\%$ of the particles to avoid boundary effects. 
Figure~\ref{slope-yprofile} shows the measurements as a function of $\overline{A}$. 
The data are only approximately collapsed, suggesting that the vertical concentration gradient is determined by a different balance than what was computed in Eq.~\ref{ourA} for the average concentration. 

In the simple scenario where particles are constantly diffusing in a gravitational field, the vertical concentration profile is exponential: $\phi(y) \propto e^{-v_s y/D}$~\cite{Chandrasekhar}. 
In the present simulation model, at the height where the concentration reaches $\phi_c$, the particles undergo few collisions so that diffusion essentially turns off (see Fig.~\ref{Phi_A}b and Ref.~\cite{Corte09}). 
Thus, we approximate the concentration profile as an exponential up to a finite height where $\phi = \phi_c$, with $\phi=0$ above that level. 
This constraint plus the conserved number of particles yields a unique profile $\phi(y)$, with a vertical concentration gradient given by:
\begin{equation}
\phi'(y)=-\frac{\pi d^2 \kappa v_s^2}{4 D^2} \frac{e^{-v_s y/D}}{1-e^{-v_s h_\infty/D}}, 
\label{gradient}
\end{equation}

\noindent where $h_\infty$ is the steady-state height of the suspended bed of particles. 
\noindent In a critical state with $h_\infty = h_c = \pi d^2 \kappa / 4 \phi_c$ and at half the bed height ($y=h_\infty/2$), the second term reduces to $e^{-2\overline{A}}/(1-e^{-4\overline{A}})$. 
This expression is order one at $\overline{A} = 1$, but it varies widely as a function of $\overline{A}$. 
Nonetheless, the first factor in Eq.~\ref{gradient} collapses the data very well, as shown in Fig.~\ref{slope-yprofile}.  
Fitting for the numerical prefactor, we find:
\begin{equation}
  \bigg|\frac{\Delta\phi}{\Delta y}\bigg| \approx 0.27 \frac{d^{2}\kappa v_s^{2}}{D^{2}}.
  \label{x_coordinate_slope}
\end{equation}

\begin{centering}
\begin{figure*}[t]
  \includegraphics[width=12.8cm]{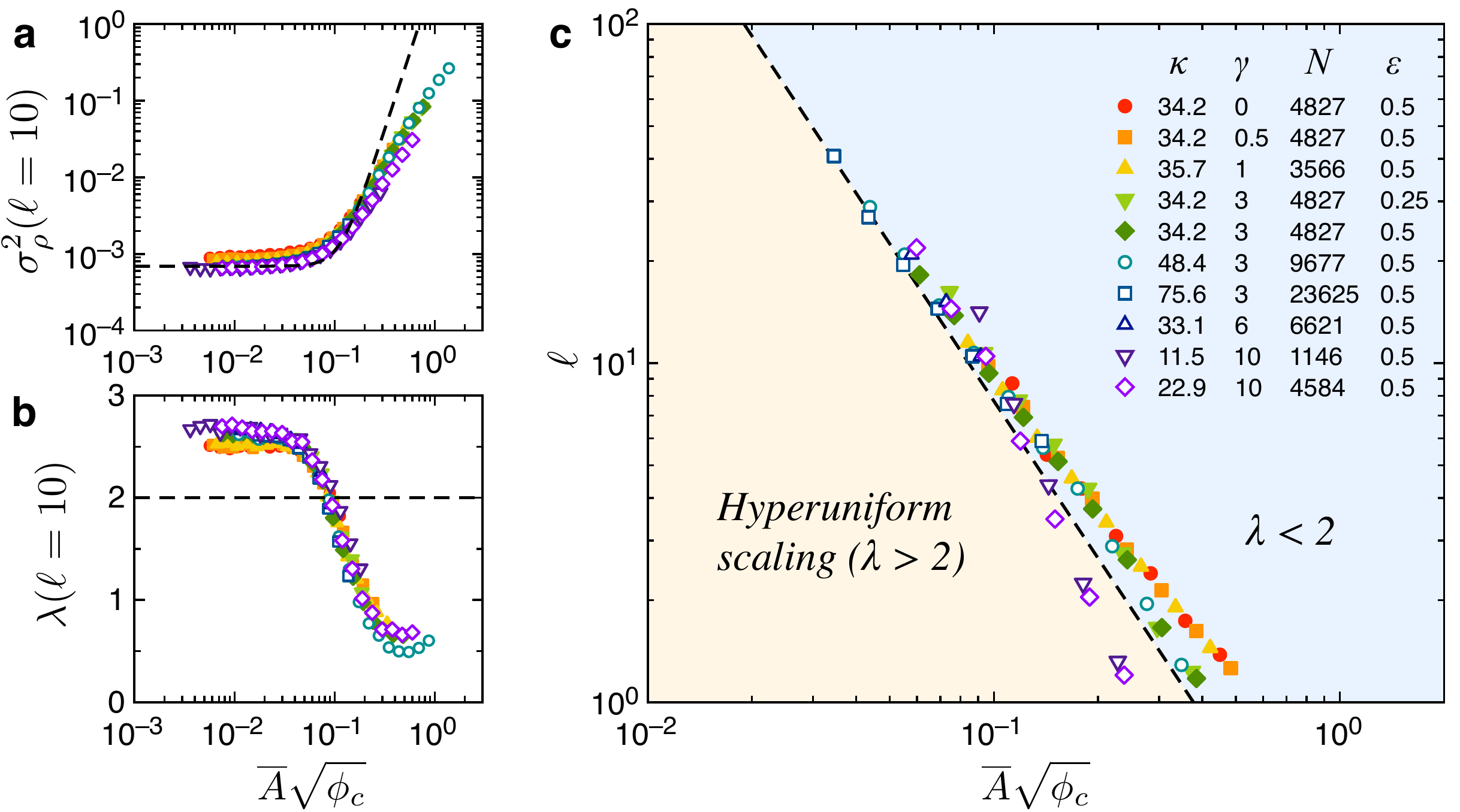}
  \caption{
  \textbf{Self-organized hyperuniformity.} 
  \textbf{(a)} 
  Variance of the number density, $\sigma_\rho^2(\ell)$, for sampling windows of size $\ell=10$. 
  The data over a wide range of parameters are collapsed when plotted versus $\overline{A}\sqrt{\phi_c}$. 
  At low velocity (i.e., low $\overline{A}\sqrt{\phi_c}$) the data plateau to the value in the critical state. 
  The data at low and moderate velocities are captured by Eq.~\ref{linear_grad} for the effect of vertical concentration gradients (dashed line). 
  \textbf{(b)}
  Magnitude of the local scaling exponent, $\lambda$, measured at $\ell = 10$. 
  At this lengthscale, hyperuniform scaling ($\lambda>2$) is observed for $\overline{A}\sqrt{\phi_c} \lesssim 0.08$. 
  \textbf{(c)} 
  Phase diagram for hyperuniform density fluctuations. 
  Hyperuniformity emerges below a finite threshold value of $\overline{A}\sqrt{\phi_c}$, and it extends to longer lengthscales as the control parameter $\overline{A}\sqrt{\phi_c}$ decreases. 
  Symbols: $\ell_\text{H}$, defined as the shortest lengthscale where the local scaling exponent $\lambda$ of the variance $\sigma_\rho^2(\ell)$ becomes shallower than 2. 
  Dashed line: Phase boundary from our theory with no free parameters, Eq.~\ref{phase_boundary}. 
  }
  \label{lambda_vs_A}
\end{figure*}
\end{centering}

\medskip
\noindent\textbf{Phase diagram.} 
We can now demonstrate how hyperuniform scaling is achieved for small concentration gradients in the full simulations. 
We insert Eq.~\ref{x_coordinate_slope} for the size of the vertical concentration gradient into Eq.~\ref{var_grad} for its effect on the variance of the number density. 
We take $\ell \ll h$ (thereby ignoring boundary effects due to the window encountering the edge of the system), and we assume a critical state where $h_\infty=h_c$. 
Plugging in and expressing in terms of $\overline{A}$, we get $\sigma_\rho^2(\ell)_\text{grad} \approx 4.1 \overline{A}^{4} \phi_c^2$, which together with Eq.~\ref{decomposition} gives: 
\begin{equation}
\sigma_\rho^2(\ell)_\text{total} \approx \sigma_\rho^2(\ell)_c + 4.1 \overline{A}^{4} \phi_c^2.
\label{linear_grad}
\end{equation}

\noindent This simple expression says that the total variance is the sum of a term from the statistics of the critical state, $\sigma_\rho^2(\ell)_c$, and a term that depends on sedimentation via the product $\overline{A}\sqrt{\phi_c}$. 
To test this result, Fig.~\ref{lambda_vs_A}a shows the variance $\sigma_\rho^2(\ell)$ measured at a lengthscale $\ell=10$ in our sedimentation simulations, as a function of $\overline{A}\sqrt{\phi_c}$. 
The data are collapsed, and they compare well with Eq.~\ref{linear_grad} up to moderate velocities. 
In Fig.~\ref{lambda_vs_A}b, we plot the magnitude of the scaling exponent, $\lambda$, measured locally at $\ell=10$. 
The data are again collapsed at low and moderate velocity, and they show hyperuniform scaling (i.e., $\lambda>2$) for sufficiently small $\overline{A}\sqrt{\phi_c}$. 

We construct a phase diagram by measuring the local scaling of $\sigma_\rho^2(\ell)$ in the same manner, as a function of $\ell$ and $\overline{A}\sqrt{\phi_c}$. 
In particular, Fig.~\ref{lambda_vs_A}c shows our measurements of the lengthscale $\ell_\text{H}$ where $\lambda$ falls below $2$, marking a phase boundary between hyperuniform and non-hyperuniform scaling. 
This lengthscale becomes larger for smaller $v_s$ (and hence smaller $\overline{A}$) for the simple reason that hyperuniform scaling is lost when density fluctuations due to the vertical concentration gradient (scaling as $\overline{A}^{4} \phi_c^2$ independent of $\ell$) become comparable to the density fluctuations in the critical state ($\sigma_\rho^2(\ell)_c \sim \ell^{-2.60}$), as anticipated by Eq.~\ref{linear_grad}. 
Equating these two terms yields the scaling: $\ell_\text{H} \sim (\overline{A}\sqrt{\phi_c})^{-4/2.60} \sim (\overline{A}\sqrt{\phi_c})^{-1.54}$. 

Improving on this scaling result, we can predict the precise location of this phase boundary by solving for the lengthscale $\ell_\text{H}$ where the local scaling exponent of Eq.~\ref{linear_grad} (i.e., $\ell$ times the logarithmic derivative of Eq.~\ref{linear_grad}) is equal to $\lambda=2$. 
This computation yields:
\begin{equation}
\ell_\text{H} \approx 0.22 (\overline{A}\sqrt{\phi_c})^{-1.54}, 
\label{phase_boundary}
\end{equation} 

\noindent which agrees very well with our data, as shown in Fig.~\ref{lambda_vs_A}c. 
We also obtain a good description of the 1D simulations by applying the same arguments in that setting (see Supplementary Note 4 and Supplementary Fig.~4). 
Three measured numerical values have entered into this calculation of the phase boundary: the scaling exponent and numerical prefactor for $\sigma_\rho^2(\ell)_c$, and the numerical prefactor for the size of the vertical concentration gradients in Eq.~\ref{x_coordinate_slope}. 
The predicted $\ell_\text{H}$ is otherwise completely constrained by our physical arguments. 

Finally, we note that all the data in Fig.~\ref{lambda_vs_A}a-c across a wide range of strain amplitude ($0<\gamma<10$) are in reasonable agreement. 
Although there are some differences for larger velocities in Fig.~\ref{lambda_vs_A}c, the data for different $\gamma$ merge together as $\overline{A}\sqrt{\phi_c}$ decreases. 
Hence, for the simulation algorithm and protocol studied here, there appears to be no practical limit on the value of $\gamma$ for preparing a hyperuniform sample. 
We have even considered the case where $\gamma=0$, in which particles receive kicks only when they come in contact with each other. 
Although this limit is not so physical, it suggests that all that is needed is for particles that are within a finite interaction region to displace each other.

\bigskip
\noindent\textbf{\large Discussion}\\
We have proposed and demonstrated a simple method for obtaining homogeneous distributions of particles in non-Brownian suspensions.  
The ingredients are extremely simple: we take advantage of a density mismatch between the particles and the fluid that is common in real settings, plus cyclic shear flow. 
This protocol could be used to ease processing demands in applications. 
Of course, there are other means for evenly distributing particles in a fluid; chaotic advection has recently been proposed as another route for homogenizing a suspension~\cite{Weijs17}. 
The key advantage of our method is that the driving amplitude does not have to be set to a specific critical value. 
More broadly, we have shown that even in the presence of body forces on the particles, local collisions are sufficient to reach and maintain a homogenous state with hyperuniform scaling. 

Looking beyond rheological behaviors, hyperuniform distributions of scattering sites can endow disordered materials with isotropic photonic band gaps~\cite{Florescu09,Man13}. 
Our method could potentially be used to prepare colloidal suspensions with such optical properties, without the need to fine-tune the driving~\cite{Hexner15}. 
Moreover, by changing the driving amplitude, the mean particle spacing can be varied continuously while maintaining a hyperuniform state. 

Surprisingly, our work has revealed three distinct combinations of the parameters $\kappa$, $v_s$, $\phi_c$, and $D$ that control self-organization in this system. 
The criteria for obtaining the critical concentration is set by the dimensionless parameter $\overline{A} \propto \kappa v_s/\phi_c D$, vertical concentration gradients scale with $\kappa v_s^2/D^2$, and hyperuniformity is controlled by $\overline{A} \sqrt{\phi_c} \propto \kappa v_s / \sqrt{\phi_c} D$. 
By considering the interplay between these effects, we have identified an emergent lengthscale $\ell_\text{H}$ beyond which hyperuniform scaling breaks down. 
This lengthscale, arising from a competition between local organization and large-scale gradients, is sufficiently general that it should arise in other settings.

\bigskip
\noindent\textbf{\large Methods}\\
\noindent\textbf{Simulations:} 
We simulate the above algorithm in domains of size $50 < L/d < 312.5$ in 2D and $750 < L/d < 7500$ in 1D. 
In the simulations with sedimentation, we begin by applying $\sim$$10^7$ shearing cycles in 2D and $\sim$$10^9$ cycles in 1D to ensure the system has reached a steady state. 
To reduce scatter in the data, we average over multiple systems. 
For the variance measurements in 2D, we also average over multiple configurations within a system by taking samples once every $10^5$ cycles in the steady state. 
Our total simulation time is approximately equivalent to 1 month of $1000$ single-core CPUs. 

\medskip
\noindent\textbf{Data availability:} 
The data that support the findings of this study are available from the corresponding author upon reasonable request.

%

\bigskip
\noindent\textbf{\large Acknowledgments}\\
We thank Nathan Keim for enlightening discussions, and we are particularly grateful to Laurent Cort\'e. 
J.M.S. acknowledges NSF-DMR-CMMT-1507938 for financial support. 
J.W. and J.D.P. gratefully acknowledge the Donors of the American Chemical Society Petroleum Research Fund for partial support of this research. 
This research was supported in part through computational resources provided by Syracuse University, including assistance from Larne Pekowsky under NSF award ACI-1541396.

\bigskip
\noindent\textbf{\large Author contributions}\\
J.D.P. conceived the study; J.W. and J.D.P. conducted simulations; J.W., J.M.S., and J.D.P. interpreted the results and wrote the paper.

\bigskip
\noindent\textbf{\large Additional information}\\
\textbf{Competing interests:} The authors declare no competing financial or non-financial interests. 

\medskip
\noindent
This is a post-peer-review, pre-copyedit version of an article published in Nature Communications. 
The final authenticated version is available at: http://dx.doi.org/10.1038/s41467-018-05195-4

\end{document}